\documentclass[aps,prb,twocolumn,showpacs,groupedaddress]{revtex4}
\usepackage{dcolumn}
\usepackage{bm}
\usepackage{graphicx}
\begin{document}

\newcommand{\azu}{Cu$_3$(CO$_3$)$_2$(OH)$_2$}

\preprint{APS/123-QED}

\title{The magnetic and crystal structure of azurite Cu$_3$(CO$_3$)$_2$(OH)$_2$ as determined by neutron diffraction}
\author{K.C. Rule$^1$, M. Reehuis$^1$, M.C.R. Gibson$^{1,2}$, B. Ouladdiaf$^3$, M.J. Gutmann$^4$, J.-U. Hoffmann$^1$, S. Gerischer$^1$, D.A. Tennant$^{1,2}$, S. S\"ullow$^5$, M. Lang$^{6}$}
\address{
$^1$Helmholtz-Zentrum Berlin, Berlin, Germany\\
$^2$Institut f\"ur Festk\"orperphysik, TU Berlin, Berlin, Germany\\
$^3$Insitut Laue-Langevin, rue Jules Horowitz, Grenoble, France \\
$^4$ISIS, Rutherford Appleton Laboratories, Chilton, UK \\
$^5$Institut f\"{u}r Physik der Kondensierten Materie, TU Braunschweig, Braunschweig, Germany\\
$^6$Goethe Universit\"{a}t, Frankfurt(M), SFB/TR 49, Germany\\} 

\begin{abstract}
Here we present neutron diffraction results on the mineral azurite. We have found that the crystal structure of azurite can be described in the space group $P2_1$ which is the next lower symmetric group of $P2_1/c$ as found in earlier works.  This small change in symmetry does not greatly influence the lattice parameters or atomic fractional coordinates which are presented here for single crystal diffraction refinements.  The ordered magnetic moment structure of this material has been determined and is comprised of two inequivalent magnetic moments on copper sites of magnitude 0.684(14) and 0.264(16) $\mu_{B}$. This result is discussed in terms of the anisotropic exchange and Dzyaloshinskii-Moriya  interactions. It is found that the system is likely governed by one-dimensional behaviour despite the long-range ordered ground state. We also highlight the significance of strain in this material which is strongly coupled to the magnetism.
\end{abstract}

\maketitle

\section{Introduction}

\begin{figure}[!ht]
\begin{center}
\includegraphics[width=0.9\linewidth]{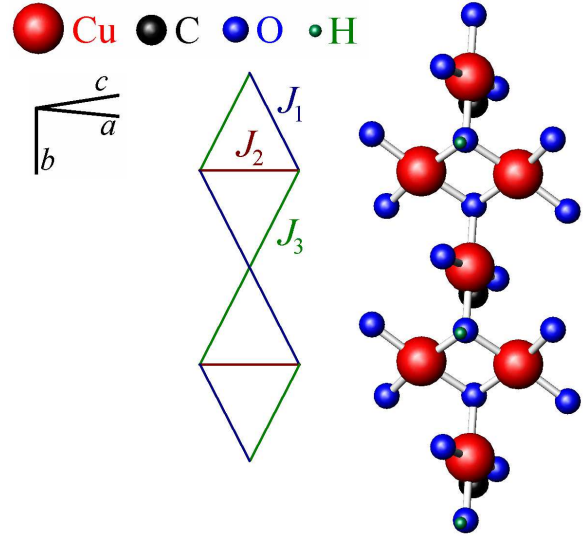}
\end{center}
\caption{(Colour online) Diamond chain model showing the relative exchange interactions.  It is now widely believed that $J_2$ is the strongest interaction in azurite.}
\label{fig:exchange}
\end{figure}

Low-dimensional, quantum spin systems have garnered much attention of late from both experimentalists and theorists alike. 
One such material is azurite, \azu, a natural mineral which has been proposed as an experimental realization of the 1D distorted diamond chain model.  
Interest in the magnetic properties of this material began in the 1950s when an ordered antiferromagnetic (AFM) ground state was observed below 1.9 K.\citep{spence,forstat} 
From these studies, a detailed description of the magnetic moment structure was never revealed however it has been speculated that the low ordering temperature may result from frustration in the diamond arrangement of spin-1/2 Cu$^{2+}$ ions\citep{kikuchi1}.  
Early susceptibility measurements indicated that the coupling of spins within the chain could be described with alternating dimers and monomers\citep{frikkee}  (Fig. \ref{fig:exchange}).  
This was later confirmed by magnetization measurements which revealed a distinct plateau at 1/3 the saturation magnetization indicating a polarization of the monomer spins within an applied magnetic field\citep{kikuchi2}.

More recently, interest in azurite has been renewed with an ongoing debate about the relative strength of the exchange interactions in this material as defined in Fig. \ref{fig:exchange} \citep{kikuchi2,gu,rule,whangbo,honecker}.  
Current consensus is that this system cannot be described in terms of a simple isotropic exchange chain Hamiltonian, but rather that inter-chain coupling and anisotropic exchange must also be taken into account when describing this material.  
A detailed understanding of the magnetic structure, as presented here, may provide further useful material in the construction of an effective Hamiltonian to describe this system.

To fully describe the magnetism in azurite, a complete understanding of the structural properties is also necessary. A definitive set of lattice parameters and atomic positions should provide an accurate input for exchange coupling calculations. 
The crystal structure of azurite was already determined in the late 1950s by Gattow and Zeeman\citep{gattow}.  It was found that azurite crystallises in the monoclinic space group $P2_1/c$ (No. 14).  This space group was confirmed by two independent studies by Zigan {\it et al.}\citep{zigan} and Belokoneva {\it et al.}\citep{belokoneva}.

More recently, evidence of magnetoelastic coupling in this material has been revealed\citep{fabris}. 
Confirmed by preliminary neutron scattering investigations, the structural strain in this material coincides with the onset of magnetic ordering\citep{gibson}. 
It is therefore possible that the structure of this material is appreciably different below $T_{\textrm{N}}$ compared with the structure published for room temperature\citep{zigan}. 
Given the interest in predicting the magnetic properties of azurite from structural considerations alone it seems prudent to obtain accurate structural data on azurite in its magnetically ordered state.   
Here we present the results of a neutron diffraction study of azurite at temperatures above and below $T_{\textrm{N}}$, revealing further evidence of structural distortion in addition to the ordered magnetic ground state. 

\section{Experimental methods}

The sample used in this study was cut from a large high-quality crystal of azurite which has been used in previous studies\citep{rule}. The sample was roughly cubic in shape (dimensions $\sim3\times4\times3$ mm) and mounted on a copper pin which offered good thermal conductivity. 
Diffraction measurements were performed on both the time-of-flight (TOF) instrument SXD at ISIS in the UK and the four-circle diffractometer D10 at Institute Laue-Langevin (ILL) in France.  
Additional diffraction measurements were also conducted using the E1 triple axis spectrometer (TAS) at Helmholtz Zentrum Berlin (HZB), Germany. 

The Single-Crystal Diffractometer SXD combines the time-sorted Laue method with a large array of position sensitive detectors to allow access to large volumes of reciprocal space in one simultaneous measurement.
Using this instrument, the sample is illuminated by a white beam of neutrons with a wavelength range 0.2-10.0\AA. A full data set of Bragg peak intensities was measured at both 1.5 and 5 K for a comparison of the structural parameters above and below the N\'eel temperature.

Additional single crystal diffraction experiments were carried out on the instrument D10.  Using a dilution $^3$He cryostat, it was possible to determine structural and magnetic properties at both 200 mK and 5 K.  This ensured that the magnetic structure of the sample was determined well below the ordering temperature of 1.87 K.  
D10 uses a Cu-monochromator selecting the single neutron wavelength $\lambda\ = 1.26$ \AA. A selection of half-integer reflections and reflections forbidden by the $P2_1/c$ space group were also measured to ensure an accurate evaluation of the higher order scattering intensity and structural symmetry respectively.  The presence of multiple scattering could be excluded by the use of psi scans.
The integration of the Bragg reflections was performed using two different methods, the SEED program\citep{peters} and the RACER program\citep{Wilkinson}, in order to optimise the quality of data.   The refinements of the crystal structure were carried out with the program Xtal 3.4 \cite{xtal}. Here the nuclear scattering lengths \textit{b}(H) =  3.7409 fm, \textit{b}(H) = 6.6484 fm, \textit{b}(O) = 5.805 fm, and \textit{b}(Cu) = 7.718 fm were used \cite{ITC}.

For the investigation of the magnetic structure of azurite, we collected data sets at D10 using the longer neutron wavelength $\lambda\ = 2.36$ \AA. We collected a set of nuclear reflections at 200mK in order to determine the overall scale factor from the crystal structure refinements. With the absorption and extinction corrected magnetic structure factors, we were able to obtain the magnetic moments of the Cu$^{2+}$-ions in the magnetically ordered range. The moments of the Cu-atoms were refined with the program FullProf \cite{fullprof}. The magnetic form factors of the Cu$^{2+}$-ion was taken from Ref. 20. 

Finally to investigate the effects of applied magnetic fields on the structure of azurite, single-crystal neutron diffraction was carried out using the E1 triple axis spectrometer at HZB (neutron wavelength $\lambda = 2.428$\AA\ ). The small azurite single crystal remained on the copper mount with a horizontal $a^* - b^*$ scattering plane, such that magnetic fields up to 14 T were applied perpendicular to $b^*$, {\it i.e.}, perpendicular to the chain direction.  In this geometry the 1/3 magnetization plateau, where the monomer spins are polarised with the applied field, ranges from 11 to 30 Tesla\citep{kikuchi2}. A temperature dependence of these applied field effects was also conducted for temperatures up to 5 K. With the aid of a $^3$He cryostat insert, base temperatures of $T \sim $ 0.5 K were attained.
 
\section{Results and analysis}

\subsection{TOF diffractometer measurements}

The data from SXD have revealed that Bragg peaks of index-type $\left(h0l\right)$ with {\it l} odd were present in the scattering profile despite being forbidden by the $P2_1/c$ space group.  This observation clearly indicates the loss of the $c$-glide plane of $P2_1/c$. 
Group-subgroup relations between the space groups showed that \azu\ crystallizes either in the noncentrosymmetric monoclinic space group $P2_1$ (No. 4) or the centrosymmetric triclinic one $P\bar{1}$ (No. 2). For $P\bar{1}$ there is no limiting condition on any $\left(hkl\right)$, but for $P2_1$ the type $\left(0k0\right)$ is restricted by \textit{k} being even. Due to the fact that reflections $\left(0k0\right)$ with \textit{k} being odd could not be observed in our experiments, it can be concluded that the correct space group is probably $P2_1$. The same limiting conditions on these reflections were observed at both 1.5 K and 5 K.  This confirms the structural origin of these peaks while indicating that azurite retains this symmetry both above and below the N\'eel temperature. 
The observation of these peaks with the TOF method is unambiguous -- the measured intensity cannot be attributed to higher-order scattering contamination, as it might with other diffraction techniques.
In contrast, the 'forbidden' reflections were actually also observed in the room temperature single crystal neutron diffraction study of Zigan {\it{et al.}}\citep{zigan} but were attributed to higher-order scattering contamination. 

The atomic fractional coordinates were refined from the SXD data containing approximately 1800 unique structural reflections at the two measured temperatures of 1.5 and 5 K using the analysis program JANA2006\citep{jana}.  In the analysis of the structural Bragg peak intensities, information from data sets collected at different orientations of the crystal are not merged resulting in a large data set and complex wavelength-dependent extinction corrections.  This, coupled with additional wavelength dependent corrections, typically gives poorer $R_F$-factors for TOF methods when compared to monochromatic techniques.  The $R_F$-factor is defined as $R_F = {\sum ||{F_o}| - |{F_c}||}/{\sum |{F_o}|}$ where ${F_o}$ and ${F_c}$ are the observed and calculated structure factors respectively.   Refinements using the $P2_1/c$ space group gave a reliability of $R_F \approx 0.080$ and compare favourably with the powder refinement as seen in Table \ref{table:powderVxtal}.  The refinement of the SXD data using the $P2_1$ space group gave unphysical anisotropic thermal parameters and as such did not give accurate positional parameters. 

The powder data of \azu\ were also re-refined using the space group $P2_1$. A comparison between the refined lattice parameters from the $P2_1/c$ space group as published earlier\citep{gibson} and the $P2_1$ space group showed that despite the change in symmetry group the lattice parameters were the same within error.  For the $P2_1$ refinement of the powder neutron diffraction data taken at 1.28 K, the lattice parameters were $a = 4.99995(11)$ \AA\, $b = 5.82256(14)$ \AA\, $c = 10.33723(19)$ \AA\ with $\beta = 92.2103(17)^{\circ}$. 
 
Using the TOF technique with SXD, it was also possible to observe and index the magnetic reflections below 1.8 K thereby confirming the propagation vector of $\textbf{k} = \left(\frac{1}{2} \frac{1}{2} \frac{1}{2}\right)$, as has been observed in neutron diffraction measurements previously\citep{gibson}.
For absolute accuracy in the structure factors and somewhat more precise atomic parameters further diffraction measurements at a continuous source were also conducted. 

\begin{center}
\begin{table*}[ht]
\caption{Positional and thermal parameters of \azu\ as obtained from the structure refinements of single-crystal (on both D10 and SXD) and powder data in the monoclinic space groups $P2_1/c$. The isotropic thermal parameters $U_{is}$ are given in units of 100 \AA$^{2}$. For the refinement several thermal parameters were constrained to be equal. In these cases the standard deviation is listed only for one of the equal parameters. }
{\small
\hfill{}
\begin{tabular}{|c|c|c|c|c|c|c|c|c|c|c|c|c|c|}
\hline
 & & \multicolumn{4}{|c|}{Powder data at 1.28 K}& \multicolumn{4}{|c|}{Single-crystal data at 5 K (SXD)}& \multicolumn{4}{|c|}{Single-crystal data at 200 mK (D10)} \\
\hline
$Atom$ & site & $x$ & $y$ & $z$ & $U_{is}$ & $x$ & $y$ & $z$ & $U_{is}$& $x$ & $y$ & $z$ & $U_{is}$\\
\cline{3-9}
\hline
Cu1&$2a$&0 &0 &0 &0.40(4)& 0 & 0 & 0 & 0.14(1)&0 &0 &0 &0.12(3) \\
Cu2&$4e$&	0.2508(6)&	0.4967(5)&	0.0834(3)&	0.40& 0.2515(3) & 0.4976(1) & 0.08324(9) & 0.25(1) &0.2516(2)&	0.4977(2)&	0.0834(1)&	0.12\\
C &	$4e$&	0.3308(6)&	0.2994(5)&	0.3192(3)&	0.63(7)& 0.3294(3) & 0.2993(2) & 0.3180(1) & 0.26(2) &	0.3303(2)&	0.2992(2)&	0.3178(1)&	0.24(3)\\
O1&	$4e$&	0.0975(7)&	0.3972(6)&	0.3318(3)&	0.60(3)&  0.1012(4)&   0.3991(2)& 0.3310(1) & 0.36(2) &	0.1016(3)&	0.3992(2)&	0.3309(2)&	0.31(3)\\
O2&	$4e$&	0.0762(7)&	0.8126(6)&	0.4451(3)&	0.60&  0.0741(4)& 0.8126(2)& 0.4450(1) &0.25(2) &	0.0743(3)&	0.8119(2)&	0.4450(2)&	0.28(4)\\
O3&	$4e$&	0.4518(6)&	0.2098(6)&	0.4183(3)&	0.60& 0.4500(4) & 0.2085(2) & 0.4176(1) & 0.35(2) &	0.4503(3)&	0.2093(2)&	0.4179(2)&	0.30(3)\\
O4&	$4e$&	0.4339(6)&	0.2949(6)&	0.2065(3)&	0.60& 0.4310(4) & 0.2937(1) & 0.2068(1) & 0.37(2) &	0.4317(3)&	0.2946(2)&	0.2072(2)&	0.37(4)\\
H	& $4e$&	0.182(1)&	0.800(1)&	0.3709(7)&	1.75(15)& 0.187(1) & 0.7996(5) & 0.3682(3) & 1.72(6) &	0.1833(7)&	0.7996(6)&	0.3686(4)&  	1.75(7)\\
\hline
\end{tabular}}
\hfill{}
\label{table:powderVxtal}
\end{table*}
\end{center}

\subsection{Crystal structure from the four-circle diffraction measurements}


The crystal structure of \azu\ was refined using the data from D10 collected at both 200 mK and 5 K.  
Refinements were carried out initially in the monoclinic space group $P2_1/c$ as done earlier by Zigan {\it et al.}\citep{zigan} and Belokoneva {\it et al.}\citep{belokoneva}.  The lattice parameters were taken from the powder data refinement since they were considered to be more accurate.

\begin{figure}[!ht]
\begin{center}
\includegraphics[width=1.0\linewidth]{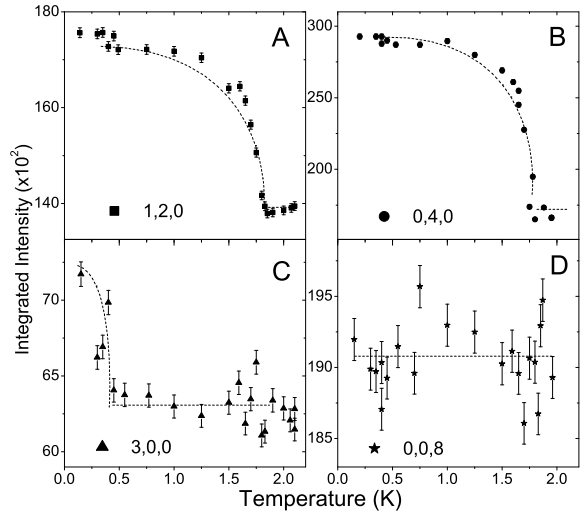}
\end{center}
\caption{The temperature dependence of the integrated intensity of the (120), (040), (008) and (300) Bragg peaks of Cu$_3$(CO$_3$)$_2$(OH)$_2$, as obtained from single-crystal neutron diffraction experiments. Note that the large increase in intensity at the N\'eel temperature is only observed for peaks with a non-zero {\it k}-index, while peaks with a non-zero {\it h}-index also exhibit a change in intensity below 0.5 K.  Lines are a guide to the eye.}
\label{fig:intensity}
\end{figure}

For the refinement of the crystal structure at 200 mK a total number of 1771 (845 unique) reflections were used. The refinement of the overall scale factor, extinction parameter and the positional and isotropic thermal parameters resulted in a poor residual $R_F$ = 0.089 ($wR_F$ = 0.110). For the extinction correction in the structural refinement the formalism of Zachariasen (type I) was used\citep{zachariasen}. The refineable parameter $g$ in this formalism is related to the mosaic distribution and assumes a Gaussian distribution of mosaic blocks within the sample.  An absorption coefficient $\mu = 0.158$ mm$^{-1}$, was also used for refinements. The refined extinction parameter ($g$) of about 1200 rad$^{-1}$ indicates that extinction is quite strong in azurite. It could be seen that strong reflections of the series $\left(h0l\right)$ were observed to be systematically weaker than the calculated values while the intensities of the $\left(0k0\right)$ reflections were calculated to be much stronger than their observations. This clearly highlights the anisotropy of the extinction. 

The significance of extinction effects in this material was revealed on measuring the temperature dependence of the Bragg peak intensities.
As discussed in previous work\citep{gibson}, such a large intensity increase below $T_{\textrm{N}}$ cannot be due to magnetic ordering alone.  
However since the intensity change coincides with the magnetic ordering transition it is clear that the structure and magnetism of azurite are closely coupled. In Fig. \ref{fig:intensity}, for example, we see a 30\% increase in the intensity of the $\left(1 2 0\right)$ peak on cooling the single crystal sample below $T_{\textrm{N}}$ and an 80\% increase in the intensity of the $\left(0 4 0\right)$. On the other hand no change of intensity was observed for the  $\left(0 0 8\right)$.  In fact only structural Bragg reflections $\left(hkl\right)$ with a large $k$ component were effected by a strong change of intensity. Surprisingly, additional yet weak intensity changes were also observed below 0.5 K on the $\left(h00\right)$ peaks suggesting that there may also be strain coupled to the {\it a}-direction.
This appears at the same temperature as an anomaly observed in recent ultrasonic measurements\citep{cong} however the nature of this anomaly is not yet known and requires further investigations.  

In order to improve the refinements we rejected 123 (58 unique) of the very strong reflections. In Table \ref{table:powderVxtal} it can be seen that the positional and thermal parameters of the different atoms could be determined with good accuracy. The refinement finally resulted in a better residual $R_F$ = 0.070 ($wR_F$ = 0.069) although the positional parameters obtained from both refinements showed a good agreement. Further, the positional parameters of the single-crystal data also show a good agreement with the values obtained from the neutron powder diffraction data collected at 1.28 K and the single-crystal data from SXD (Table \ref{table:powderVxtal}). But despite the fact that the refinement quality from the single-crystal data was reduced due to anisotropic extinction we were able to determine the positional parameters with much better accuracy than from the powder diffraction study (Table \ref{table:powderVxtal}). The refinements of the D10 single-crystal data collected at 5 K resulted in a strongly enlarged extinction parameter of about 4000 rad$^{-1}$ which is clearly also affecting the fit parameters of the SXD data. 

\begin{figure}[!ht]
\begin{center}
\includegraphics[width=0.9\linewidth]{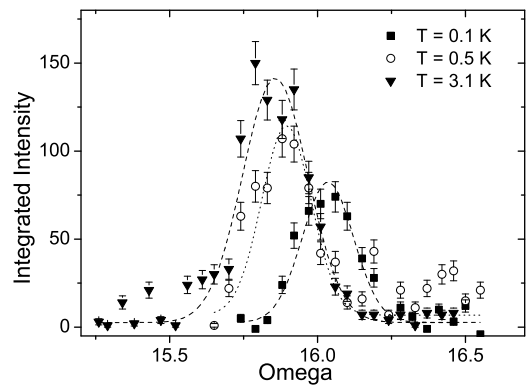}
\end{center}
\vspace{-3.5em}
\caption{Intensity profile of the [101] Bragg reflection at temperatures above and below $T_{\textrm{N}}$.  This reflection is forbidden  for the $P2_1/c$ space group but allowed for $P2_1$.  Lines are guides to the eyes.}
\vspace{-0.5em} 
\label{fig:forbidden}
\end{figure}

\begin{figure}[!ht]
\begin{center}
\includegraphics[width=0.7\linewidth]{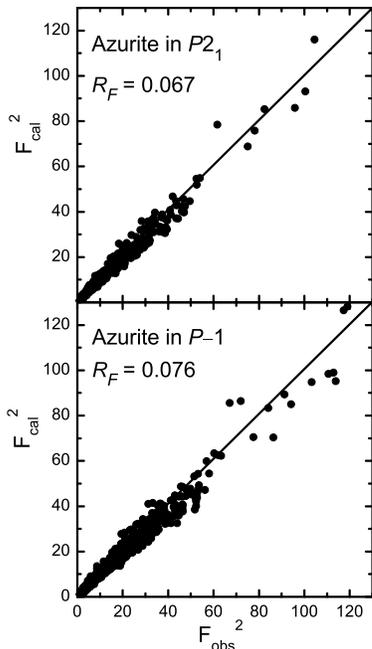}
\end{center}
\vspace{-3.5em}
\caption{Comparison of observed and calculated structure factors (in $F^2$) for the $P2_1$ (upper) and $P\bar{1}$ (lower) space groups.  The distribution of points around the line $F_{obs}^2$ = $F_{calc}^2$ indicates the high quality of the fits.  These data have been corrected for extinction effects as mentioned in the text.}
\vspace{-1.0em} 
\label{fig:FobsFcalc}
\end{figure}

It was again found in the D10-experiment that reflections of the type $\left(h0l\right)$ with \textit{l} being odd had weak yet significant intensity. The intensity profiles of the reflection $\left(101\right)$ at different temperatures is presented in Fig. \ref{fig:forbidden}. Finally a total of 60 unique reflections of this type could be collected. The presence of multiple scattering was checked by the use of psi-scans. Since the reflections of the type $\left(0k0\right)$ with $k$ = odd were again not observed, it is likely that the crystal structure of azurite can be described in the space group $P2_1$. However we carried out the structure refinements using both the monoclinic and triclinic space groups $P2_1$ and $P\bar{1}$, respectively. The refinement of the single crystal data gave a residual $R_F$ = 0.067 ($wR_F$ = 0.063).  A significantly larger residual $R_F$ = 0.076 ($wR_F$ = 0.074) was obtained from the refinement of 51 parameters using the monoclinic space group $P\bar{1}$. A comparison of observed and calculated structure factors (in $F^2$) is shown in Fig. \ref{fig:FobsFcalc}. It was interesting to see that the calculated intensities of the reflections $\left(010\right)$, $\left(030\right)$, $\left(050\right)$, $\left(070\right)$ [$\left(0k0\right)$ with {\it k} = odd, forbidden in $P2_1$)] were practically equal to zero. This also indicates that the structure of azurite still contains the screw axis $2_1$ parallel to \textit{b}. Therefore we can conclude that azurite crystallizes in the monoclinic space groups $P2_1$. The results of the refinement of the D10 data for the $P2_1$ symmetry group are summarised in Table \ref{tab:FractCoords}. 

The positional parameters of the copper atoms for the two different space groups $P2_1/c$ and $P2_1$ are given in Tables \ref{table:powderVxtal} and \ref{tab:FractCoords}. In $P2_1/c$ the Cu1-atoms are located at the special Wyckoff position $2a(0, 0, 0)$. Therefore in this space group the interatomic distances to the next neighboring Cu1'-atoms in $\left(0, \frac{1}{2}, \frac{1}{2}\right)$, $\left(0, -\frac{1}{2}, \frac{1}{2}\right)$, $\left(0, \frac{1}{2}, -\frac{1}{2}\right)$, and $\left(0, -\frac{1}{2}, -\frac{1}{2}\right)$ are identical [$d_{Cu1-Cu1'} = 5.9321(1)$ \AA\ ]. In the space group $P2_1$ the Cu1-atoms are located at the general Wyckoff position $2a(x, y, z)$. In Table \ref{tab:FractCoords} it can be seen that the $x$-parameter does not show any change from the ideal position $x = 0$. On the other hand the $y$- and $z$- values show a slight shift from the ideal position $y = 0$ and $z = \frac{1}{4}$ (in $P2_1$) or $z  = 0$ (in $P2_1/c$) of about $3-4\sigma$. In the lower symmetric space group $P2_1$ one finds two different interatomic distances. The distances between Cu1 in (0, 0, 0) and the Cu1'-atoms in $\left(0, \frac{1}{2}, \frac{1}{2}\right)$, and $\left(0, -\frac{1}{2}, \frac{1}{2}\right)$ are the same [$d_1 = 5.913(4)$ \AA\ ]. But along the opposite $z$-direction the distance to the Cu1'-atoms in $\left(0, \frac{1}{2}, -\frac{1}{2}\right)$, and $\left(0, -\frac{1}{2}, -\frac{1}{2}\right)$ is $d_2 = 5.952(4)$ \AA\ where $d_1$ and $d_2$ are plotted in the $bc$-plane in Fig. \ref{fig:magstruc}. In fact our study shows that the shift from the ideal position is relatively weak. However it is interesting to note that the lower symmetric setting allows a dimerization of monomer Cu1-sites. On the other hand the Cu2-atoms (in $P2_1$/c) are located in the general Wyckoff position $4e(x, y, z)$, while in the lower symmetric space group $P2_1$ this position splits into two different positions Cu21 and Cu22, but both at the general position $2a(x, y, z)$. In Tables \ref{table:powderVxtal} and \ref{tab:FractCoords} it can be seen that the Cu2-atoms in the two representations show a good agreement, with respect to their standard deviations. Please note that one finds the relation $z'{\textrm{(Cu2)}} = z{\textrm{(Cu21 or Cu22)}}-\frac{1}{4}$ for the two different representations.

\begin{table}
\caption{Positional and thermal parameters of \azu\ as obtained from the structure refinements of single-crystal data taken at 200 mK using the monoclinic space group $P2_1$. For the refinement several thermal parameters, $U_{is}$ (given in 100 \AA$^{2}$), were constrained to be equal. In these cases the standard deviation is listed only for one of the equal parameters. } 
\begin{tabular}{|c|c|c|c|c|c|}
\hline
  Atom &site& $x$ & $y$ & $z$ & $U_{is}$ \\
\colrule
Cu1&	$2a$&	0.0000(5)&	0.0015(6)&	0.2511(3)&	0.13(4)\\
Cu21&	$2a$&	0.2514(4)&	0.4985(5)&	0.3336(2)&	0.16(3)\\
Cu22&	$2a$&	0.7488(4)&	0.5027(5)&	0.1668(3)&	0.16\\
C11&	$2a$&	0.3306(5)&	0.3012(5)&	0.5685(3)&	0.21(3)\\
C12&	$2a$&	0.6691(5)&	0.7026(5)&	0.9334(3)&	0.21\\
O11&	$2a$&	0.0997(5)&	0.3921(5)&	0.5815(4)&	0.27(3)\\
O12&	$2a$&	0.8954(5)&	0.5949(5)&	0.9198(4)&	0.27\\
O21&	$2a$&	0.0762(5)&	0.8100(5)&	0.6957(4)&	0.33(4)\\
O22&	$2a$&	0.9269(6)&	0.1853(5)&	0.8065(4)&	0.33\\
O31&	$2a$&	0.4518(5)&	0.2068(6)&	0.6700(4)&	0.33(3)\\
O32&	$2a$&	0.5518(5)&	0.7891(6)&	0.8341(4)&	0.33\\
O41&	$2a$&	0.4366(5)&	0.2915(5)&	0.4564(4)&	0.34(3)\\
O42&	$2a$&	0.5732(5)&	0.7037(5)&	0.0411(4)&	0.34\\
H11&	$2a$&	0.1768(12)&	0.8088(11)&	0.6129(7)&	1.60(7)\\
H12&	$2a$&	0.8117(12)&	0.2113(23)&	0.8754(7)&	1.60(7)\\
\colrule \colrule
\end{tabular}
\label{tab:FractCoords}
\end{table}

\subsection{High magnetic field measurements}

\begin{figure}[!ht]
\vspace{-1.0em}
\begin{center}
\includegraphics[width=1.0\linewidth]{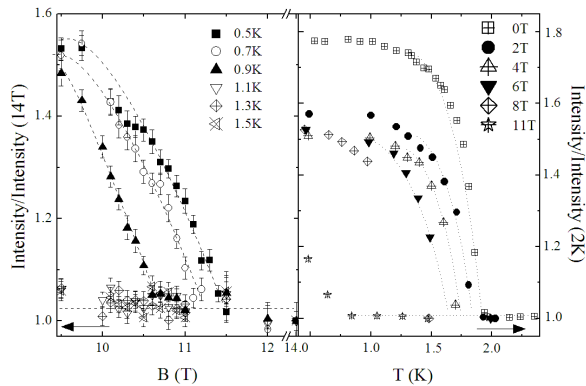}
\end{center}
\vspace{-2.5em}
\caption{(left) The field (left) and temperature (right) dependence of the normalized
integrated intensity of the $\left(120\right)$ Bragg peak of
\azu\ near the plateau field range from single crystal
neutron diffraction.  Lines are guides to the eye.}
\label{fig:highfield}
\end{figure}

Further insight into the magnetic and structural interdependence can be taken from the field dependence of the intensity of the $\left(1 2 0\right)$ structural Bragg peak.  For this, measurements were conducted using the E1 spectrometer at HZB.
The left panel of Fig. \ref{fig:highfield} displays the integrated intensity of the $\left(1 2 0\right)$ nuclear Bragg reflection as a function of field while the right panel shows the temperature dependence. The intensity has been normalised either to the peak intensity at 14 T or at 2 K, since no further intensity changes were observed well within the plateau phase and above the ordering temperature respectively.  At temperatures approaching $T_{\textrm{N}}$, the field at which the intensity becomes invariant is reduced.  

The atomic fractional coordinates were not observed to change significantly within the accuracy of these neutron scattering experiments.  However field and temperature induced changes in the lattice parameters have been observed in more accurate studies\citep{fabris, gibson}.  These slight changes may facilitate the formation of the mosaic blocks within the crystal which in turn become misaligned on a macroscopic scale increasing the mosaicity of the sample.  

\subsection{Magnetic structure}

The magnetic order of the Cu$^{2+}$-ions sets in at the N\'eel temperature  $T_{\textrm{N}}$ = 1.87 K. Previous diffraction experiments showed that the magnetic cell of azurite could be described with the propagation vector $k = \left(\frac{1}{2} \frac{1}{2} \frac{1}{2}\right)$ \cite{gibson}. In order to determine the magnetic structure of the Cu-sites we performed a symmetry analysis for the space groups $P2_1/c$ and $P2_1$. This allowed us to deduce the possible spin configurations compatible with the symmetry of the crystal structure according to the irreducible representations (ireps)  $\Gamma_\nu$ as described elsewhere \cite{bertaut}. In a magnetic structure the magnetic moment $\mu$ of an atom $j$ at the lattice point $\textbf{R}_n$ in the unit cell is characterized by the propagation vector $k$, where the moment is given by the Fourier series $\mu_{nj} = \sum_{k}\textbf{S}_{\textbf{k}j} \cdot e^{-2\pi i\textbf{k}\cdot \textbf{R}n}$. The Fourier series $\textbf{S}_{\textbf{k}j}$ are linear combinations of the basis functions of the ireps obtained from the symmetry analysis. 
In the monoclinic space group $P2_1/c$ the two different copper atoms are located (within the unit cell) at the Wyckoff position $2a$[Cu11$\left(0,0,0\right)$,  Cu12$\left(0,\frac{1}{2},\frac{1}{2}\right)$] and $4e$[Cu21$\left(x,y,z \right)$, Cu22$\left(1-x, y+\frac{1}{2}, z\right)$, Cu23$\left(1-x, 1-y, 1-z\right)$, and Cu24$\left(x, -y+\frac{1}{2}, z+\frac{1}{2}\right)$], with $x$ = 0.2513(3), $y$ = 0.4976(2), $z$ = 0.0834(2). There are four symmetry operators in $P2_1/c$ leading to four one-dimensional complex irreducible representations namely $\Gamma_1$, $\Gamma_2$, $\Gamma_3$ and $\Gamma_4$.
Since the propagation vector is $\textbf{k} = \left(\frac{1}{2} \frac{1}{2} \frac{1}{2}\right)$ then $\textbf{k}$ and $\textbf{-k}$ are equivalent such that the Fourier series $\textbf{S}_{\textbf{k}j}$ should be real. The induced representation of the $2a$ site can be decomposed as  $\Gamma_{2a} =3\Gamma_1 +3\Gamma_3$.  
The Fourier coefficients for the $2a$ site are then a linear combination of the irreducible representation $\Gamma_1$ and its complex conjugate $\Gamma_3$ i.e. $(1+i) \Gamma_1 +(1-i) \Gamma_3 $ leading to the following magnetic configuration with real Fourier components, $\textbf{S}_{\textbf{k}11} = (u, v, w)$ for the atom Cu11 and $\textbf{S}_{\textbf{k}12} = (u, -v, w)$ for the atom Cu12.
The induced representation of the $4e$ site can be written as $\Gamma_{4e} =3\Gamma_1 +3\Gamma_2  + 3\Gamma_3 +3\Gamma_4$. In a similar way we obtain the Fourier coefficients for the $4e$.
In the lower symmetric space group $P2_1$, Cu1 does not change its multiplicity and it is located at the position $2a$[Cu11$\left(x,y,z\right)$ and Cu12$\left(1-x,y+\frac{1}{2},1-z\right)$] with $x$ = 0.0000(3), $y$ = 0.0015(6), $z$ = 0.2511(3). For the atoms Cu11 and Cu12 the Fourier components are found to be the same for both space groups $P2_1$ and $P2_1/c$. Due to the fact that the centre of symmetry is lost in $P2_1$ the four Cu2 atoms split in two different sites, both in the same Wyckoff position $2a$ as also found for Cu1. Within the unit cell the Cu2 atom are located in: Cu21$\left(x,y,z \right)$, Cu22$\left(1-x, y+\frac{1}{2}, 1-z\right)$ with $x$ = 0.2514(4), $y$ = 0.4985(5), $z$ = 0.3336(2); Cu23$\left(x,y,z \right)$ and Cu24$\left(1-x, y-\frac{1}{2},1-z\right)$ with $x$ = 0.7488(4), $y$ = 0.5027(5), $z$ = 0.1668(3). 
The only difference between the different setting is that the components $u, v, w$ of the subsets (Cu21,Cu22) and (Cu23,Cu24) must not be necessarily the same. However, our crystal structure refinements showed that the positional parameters are only slightly changed in the lower symmetric space group $P2_1$. Therefore the values of $u, v, w$ were constrained to be equal for all the four Cu2-atoms. Further it is interesting to see that coefficients for the atoms Cu12, Cu22, and Cu24 are purely imaginary. Then a phase factor of $\frac{1}{4}$(in fractions of $2\pi$) between the atoms of one subset has to be taken into account.

For the refinements of the magnetic structure we used a total number of 40 independent magnetic reflections. It could be shown that the coupling between the Cu-atoms of the monomers (Cu11 and Cu12) is compatible with spin arrangement of irep $\Gamma_1$: $\textbf{S}_{\textbf{k}11} = (u, v, w)$ and $\textbf{S}_{\textbf{k}12} = (u, -v, w)$.  The moments are coupled parallel in the monoclinic $ac$-plane, and antiparallel along the $b$-axis. For subsets of the Cu2-site we found the spin configurations: $\textbf{S}_{\textbf{k}21} = (u, v, w), \textbf{S}_{\textbf{k}22} = (u, v, w), \textbf{S}_{\textbf{k}23} = (-u, -v, -w), \textbf{S}_{\textbf{k}24} = (-u, -v, -w)$. The spin coupling between Cu21 and Cu22, as well as the spin coupling between Cu23 and Cu24 along the monoclinic $b$-axis is compatible with the symmetry analysis. 

\begin{figure}[!ht]
\begin{center}
\includegraphics[width=1.0\linewidth]{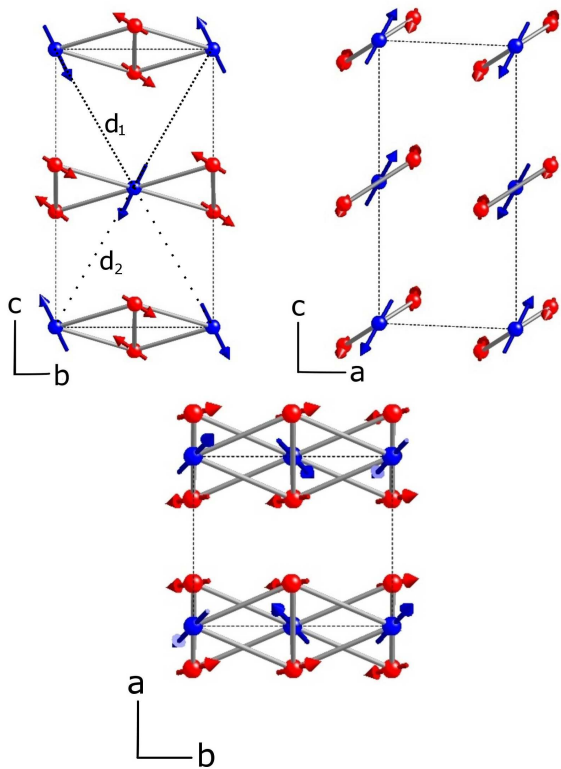}
\end{center}
\vspace{-1.5em}
\caption{(Colour online) The magnetic structure of azurite, shown in the \textit{b-c}, \textit{a-c} and \textit{a-b} planes (l.-r.). The monomer sites, with a moment size of 0.684 $\mu_{B}$, are represented by the blue (dark) spheres, while the dimer sites (0.264 $\mu_{B}$) are represented by the red (light). The monomer spins on alternate $\it{a-b}$ planes are aligned perpendicularly. The lengths $d_1$ and $d_2$ have been determined for the $P2_1$ space group in section C.}
\label{fig:magstruc}
\end{figure}

The refined magnetic structure, resulted in a residual $R_F = 0.069$. Fig. \ref{fig:magstruc} shows the refined magnetic structure where the blue (dark) spins represent the Cu(1) sites which we commonly refer to as the monomer sites and the red (light) spins represent the Cu(2) sites which form dimerised pairs.  Along the chain direction, the monomer spins are collinear and oriented in the {\it ac}-plane with a slight canting along the {\it b}-direction.  The monomer spins lie with an easy axis at an angle of $55 \pm 3^{\circ}$ from the {\it c}-axes which corresponds within error to the value of $52^{\circ}$ found from magnetisation measurements\citep{frikkee}.

The magnetic moment on the monomer sites, is 0.684(14) $\mu_{B}$, while the moment on the dimer sites is 0.264(16) $\mu_{B}$.  The relative $x$, $y$ and $z$ components of each site are shown in Table \ref{tab:MagSpin}.
A moment size of 0.684 $\mu_{B}$ is consistent with magnetic moments found in other cuprate systems where the reduction from the 1 $\mu_{B}$ expected for a Cu$^{2+}$ ion can be attributed to zero point spin fluctuations and/or covalency effects\citep{yang, tranquada}.
Note that the size of the moment on the dimers determined in this study, while significantly reduced from the monomer value, is somewhat larger than the moment found for azurite in its plateau phase from NMR\citep{aimo}.  This reduction may be caused by mean field effects. In the plateau phase, the monomers form a ferromagnetic chain which may influence the overall moment at the dimer sites.  In low fields, when the monomers are coupled antiferromagnetically, the mean field effects are somewhat different.

\begin{table}
\caption{Table of the magnetic spin components of the monomer (Cu(1)) and dimer (Cu(2)) spins at the atomic fractional coordinate positions of Cu(1)  and Cu(2).}
\begin{ruledtabular}
\begin{tabular}{ccc|cc}
&Site & Spin-component ($\mu_{B}$) & Full moment ($\mu_{B}$)&\\
\hline
   & Cu(1)$_{x}$   & 0.42 (2)&  &\\  
   & Cu(1)$_{y}$   & 0.36 (3)&  &\\ 
   & Cu(1)$_{z}$   & 0.42 (3)&  0.684 (14) &\\ 
   & Cu(2)$_{x}$   & 0.036 (17)&  &\\   
   & Cu(2)$_{y}$  & 0.243 (15) & &\\  
   & Cu(2)$_{z}$   & 0.099 (16)& 0.264 (16)&\\  
\end{tabular}
\end{ruledtabular}
\vspace{-2.0em} 
\label{tab:MagSpin}
\end{table}

All magnetic Bragg peaks showed an increase in intensity below the ordering temperature as expected for a second order phase transition. 
Since the influence of extinction is roughly proportional to the intensity of the Bragg peaks, the effect of the anisotropic extinction on the overall magnetic refinement result is minimal.

\section{Discussion}

\subsection{Extinction}

\begin{figure}[!ht]
\begin{center}
\includegraphics[width=0.95\linewidth]{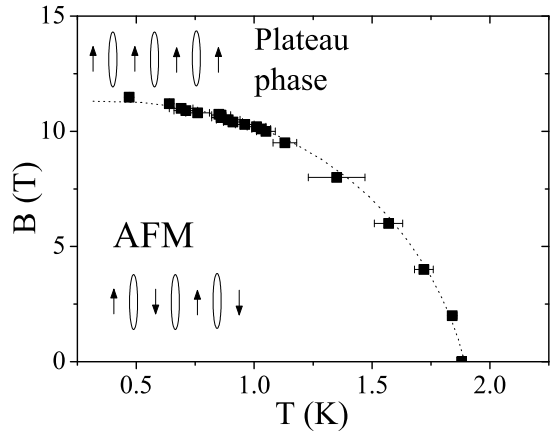}
\end{center}
\vspace{-3.5em}
\caption{An indirect measure of the phase diagram as inferred from the field and temperature dependence of the integrated intensity of the $\left(1 2 0\right)$ peak (from Fig. \ref{fig:highfield}).  Line is a guide to the eye. Ovals and arrows represent the Cu-dimer pairs and the relative orientation of the Cu-monomers repsectively.}
\label{fig:phase}
\end{figure}

Due to the close coupling between the extinction and the onset of magnetic order, it is likely that the magnetoelastic coupling is mediated by the diamond chain units which propagate along the crystallographic {\it b}-axis parallel to $\left(0 k 0\right)$.  
As a magnetic field is applied perpendicular to the {\it b}-axis, the AFM correlations between monomer spins gives way to ferromagnetic correlations which in turn allows the lattice to return to a state with fewer mosaic block discontinuities and therefore increased extinction effects.  At temperatures closer to $T_{\textrm{N}}$, the magnetic interactions are weaker and thus a lower applied field is required to reduce the spread of mosaic blocks.  This may explain why we see a reduction in intensity of the nuclear $\left(1 2 0\right)$ Bragg peak in applied fields as seen in Fig. \ref{fig:phase}.  Thus the change in intensity with applied field, whilst being an extinction effect, highlights the phase boundary of the AFM state in azurite.  It is also worthwhile to note that the intensity of the $\left(1 2 0\right)$ Bragg peak in the plateau phase is the same as in the paramagnetic, low field phase indicating that the mosaicity in both phases is the same.

\subsection{Magnetism}

For a reduced moment size of 0.264 $\mu_{B}$ we can consider that the Cu(2) sites (represented by the red spins in Fig. \ref{fig:magstruc}) are coupled primarily into spin singlet states with their nearest neighbouring Cu(2) spin in the diamond unit. The observation of a finite spin polarisation implies that the singlet state is subject to some perturbative effect which induces the non-negligible spin moment. Assuming the interchain coupling is negligible in comparison with the intrachain coupling, the staggered field of the neighbouring monomer spin sites may be responsible for this perturbation. 

Considering at first only Heisenberg exchange couplings, the expectation values of the spin moment on the dimer atoms in the staggered  field of the neighbouring monomer sites can be calculated from the Hamiltonian
\begin{equation}
\mathcal{H}=J_{2}\vec{S}_{1}.\vec{S}_{2}+(J_{1}-J_{3})\left\langle S_{m}^{z}\right\rangle S_{1}^{z}
-(J_{1}-J_{3})\left\langle S_{m}^{z}\right\rangle S_{2}^{z} 
\end{equation}

where the ordered moment on the monomer sites defines the \textit{z}-axis and is of magnitude $\left\langle S \right\rangle$. For all nonzero $\left\langle S \right\rangle$ a staggered moment on sites $S_{1}$ and $S_{2}$ is established for the ground state collinear with the monomer spins. Whilst the ratio of $\left(J_{1}-J_{3}\right)/J_{2}=1/2.8$ implied by the relative ordered moments on dimer and monomer spins is in rather good agreement with the exchange couplings determined by Rule \textit{et al.}\citep{rule} the non-collinearity requires the introduction of other exchange terms.   

\begin{figure}[!ht]
\begin{center}
\includegraphics[width=0.35\linewidth]{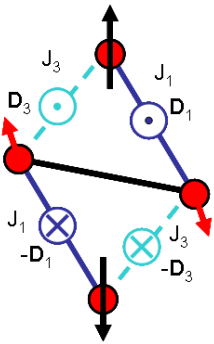}
\end{center}
\vspace{-1.5em}
\caption{(Colour online) Dzyaloshinskii-Moriya (DM)  interactions on the diamond shape units where the vertical direction corresponds to the \textit{b}-axis.  We define the notation of $\odot$ and $\otimes$ to represent antiparallel vectors since the actual orientation of the DM vector is not known.}
\label{fig:DM}
\end{figure}

A likely scenario involves Dzyaloshinskii-Moriya (DM) interactions which strongly influence the non centrosymmetric $J_{1}$ and $J_{3}$ exchange interactions as outlined in Figure \ref{fig:DM}.  Taking the extra exchange terms of the type $\vec{D}_{1}.\left(\vec{S}_{m1}\times\vec{S}_{1}+\vec{S}_{m2}\times\vec{S}_{2} \right)+\vec{D}_{3}.\left(\vec{S}_{m2}\times\vec{S}_{1}+\vec{S}_{m1}\times\vec{S}_{2} \right)$ and using the fact that $\vec{S}_{m1}=-\vec{S}_{m2}=(0,0,\left\langle{S}_{m}\right\rangle)$ then the Hamiltonian becomes 
\begin{eqnarray*}
\mathcal{H} = J_{2}\vec{S}_{1}.\vec{S}_{2} & + &(D_{1}^{y}-D_{3}^{y})\left\langle S_{m}^{z}\right\rangle (S_{1}^{x}-S_{2}^{x})\\
& - & (D_{1}^{x}-D_{3}^{x})\left\langle S_{m}^{z}\right\rangle (S_{1}^{y}-S_{2}^{y})\\
& + & (J_{1}-J_{3})\left\langle S_{m}^{z}\right\rangle (S_{1}^{z}-S_{2}^{z})
\end{eqnarray*}
This then causes a titling of the dimer spins compared to the monomer \textit{z} direction along the new effective staggered field   $(D_{1}^{y}-D_{3}^{y}, D_{1}^{x}-D_{3}^{x}, J_{1}-J_{3})\left\langle S_{m}^{z}\right\rangle$ direction. The significant tilting observed would imply that the DM interactions are of order $(J_{1}-J_{3})$, \textit{i.e.} a few K.  

A further consequence of including DM interactions is that the effective Hamiltonian of the monomer spins loses full rotational symmetry and results in an XXZ type Hamiltonian with planar character (the easy plane being perpendicular to $(D_{1}-D_{3})$). To explain the spin flop type behaviour observed in azurite\citep{Love} other terms resulting in an easy axis are needed, either from the symmetry lowering to $P2_1$, interchain coupling, or further anisotropies. The application of non-commuting magnetic fields to such low dimensional XYZ antiferromagnets, where $J_{x} \not= J_{y} \not= J_{z}$, have been observed to cause remarkable behaviour in the vicinity of quantum critical points \citep{kenzelmann, coldea}  and azurite may well provide an important new system in this line. Here, for example, the quantum phase transition into the plateau phase could take on a transverse Ising character \citep{kenzelmann}. We defer further discussion of the full derivation of the Hamiltonian to a future paper on the dynamics in azurite. 

The question remains as to what extent the diamond chains in azurite can be considered to be modeled reasonably as a one-dimensional system.  The Mermin-Wagner theorem indicates that truly one-dimensional systems cannot enter a long ranged ordered state due to quantum fluctuations\citep{mermin}, however experimentally, materials such as azurite which display quasi-1D characteristics clearly enter a 3D N\'eel state. 
The significance of quantum fluctuations in a given system may be established from the ratio of the Curie-Weiss temperature to the N\'eel temperature.
It has been found from susceptibility measurements that for azurite this value is roughly $-\Theta_{CW}/T_N \approx 5$, implying that quantum fluctuations should play a significant role for all temperatures below $\Theta_{CW}$\citep{ramirez}. 

Further, the relative strengths of inter- and intra-chain couplings may be inferred from the ordered magnetic moment for comparatively weakly coupled chains \citep{schulz, lake}. The effective one-dimensional chain model provides a reasonable description of the system with coupling $J_{mono} = 0.87$ meV, determined previously from inelastic neutron scattering\citep{rule}. Taking the measured moment of $m_{0} = 0.684 \mu_{B}$, then an effective tetragonal interchain coupling of $|J_{\perp}| = 0.097$ meV would be required to account for the ordering strength. It should be noted that Ising anisotropies would serve to increase the ordered moment for the same interchain coupling.

\section{Conclusion}

In summary, we have observed that azurite belongs to the lower symmetry space group $P2_1$.  We have revealed the significance of magnetoelastic strain in this material as observed by field and temperature dependent extinction effects. 
Also presented here is the ground state magnetic structure of azurite which confirms that the diamond chain arrangement of magnetic Cu$^{2+}$ sites in this material can, to some extent, be considered as an alternating arrangement of dimer and monomer entities.
The ordered magnetic structure may also indicate the presence of competing interactions between the chains in this material and the significance of anisotropic exchange.
Comparison of the magnitude of the ordered magnetic moment with existing theories of coupled quantum spin chains implies that the interchain coupling is weak in this material and that the system may thus be considered one-dimensional. The magnitude and orientation of the magnetic moments in azurite points to additional anisotropy terms in the Hamiltonian.
Additional diffraction studies will be required to determine the ordered magnetic state of this intriguing material in its plateau phase.

\begin{acknowledgments}
We would like to thank Wolfgang Jauch for valuable discussions.  The authors are grateful for the local support staff at ISIS, HZB and the ILL. This research was supported in part by the Deutsche Forschungsgemeinschaft DFG under Grant No. SU229/9-1.
\end{acknowledgments}

\end{document}